\documentclass[pra,twocolumn,amsmath,amssymb,aps,superscriptaddress,showpacs]{revtex4}
\usepackage{amssymb}
\usepackage{amsmath}
\usepackage{latexsym}

\usepackage{graphicx}

\def\Tr{\mathop{\rm Tr}\nolimits}

\def\sgn{\mathop{\rm sgn}}
\def\mod{\mathop{\rm Mod}}

\newtheorem{thm}{Theorem}

\begin{document}

\title{Fourier Analytic Approach to Phase Estimation}

\author{Hiroshi Imai}\affiliation{National Institute of Informatics,
2-1-2 Hitotsubashi, Chiyoda-ku, Tokyo 101-8430, Japan}
\author{Masahito Hayashi}\affiliation{Graduate School of Information Sciences, Tohoku University, Sendai, 980-8579, Japan}
\begin{abstract}
For a unified analysis on the phase estimation, we focus on the limiting distribution.
It is shown that the limiting distribution can be 
given by the absolute square of the Fourier transform of
$L^2$ function whose support belongs to $[-1,1]$.
Using this relation, we
study the relation between the variance of the limiting distribution
and its tail probability. 
As our result, we prove that the protocol
minimizing the asymptotic variance does not minimize the tail probability.
Depending on the width of interval,
we derive the estimation protocol minimizing the tail probability out of a given
interval.
Such an optimal protocol is given by a prolate spheroidal wave function 
which often appears in wavelet or time-limited Fourier analysis. 
Also, the minimum confidence interval 
is derived with the framework of interval estimation that
assures a given confidence coefficient.
\end{abstract}
\maketitle

\section{Introduction}\label{s1}
Estimating/identifying the unknown unitary operator is discussed in
both research fields of quantum computation \cite{CEMM,Kitaev} and quantum statistical
inference, therefore, it is a fundamental topic in quantum information.
In quantum computation, the unknown unitary operator is given as an
oracle, and it is discussed how many applications are required for
identifying the given unknown oracle with a given precision.
In quantum statistical
inference, in contrast, many researchers optimize the average fidelity
or mean square error between the true unitary and the obtained guess.
In the both research sides, we optimize the state inputting the unknown
unitary as well as the measurement, and quadratic speedup is reported
from the both research sides \cite{Giovannetti,Buzek,Hayashi,Imai}.
However, the both areas discuss the same topic based on the different
criterion in this manner, and there is an example such that the quadratic
speedup appears depending on the choice of the criterion. Therefore,
since the relation between both criteria is not clear, it is required
to treat this problem with a common framework.

In the present paper, as the most typical example, we focus on the
phase estimation, in which, the quadratic speedup was demonstrated
experimentally \cite{Heisenberg-limit}.
Kitaev first treated the phase estimation problem in quantum computation viewpoint
\cite{Kitaev2}.
Since it appears in Shor's
factorization, it is considered as a fundamental topic in quantum
computation as well as physics. In order to treat the quadratic speedup
more deeply, we focus on the limiting distribution, which provides
the stochastic behaviour of the estimate around the neighborhood of the true parameter. 
That is, it provides the distribution of the random variable $n (\hat{\theta}-\theta)$
when the estimate, the true parameter, and the number of applications are given as
$\hat{\theta}$, $\theta$, and $n$.
While the limiting distribution is a common concept
in statistics and was studied in estimation of quantum state \cite{GK,GJK,GJ},
it was not studied in estimation of the unknown unitary operator systematically.

The concept of `limiting distribution' is very useful for the following four points.
First, 
the variance of the limiting distribution gives
the asymptotic first order coefficient of 
the mean square error.
Second, 
the tail probability of the limiting distribution 
for a given interval provides the tail probability of the interval
when the width of the interval behaves the order $\frac{1}{n}$.
Third, 
using the limiting distribution, we can discuss the phase estimation under the framework of
interval estimation, whose meaning is explained later.
Forth, 
using this concept, we can treat the number of applications required
for attaining the given accuracy, i.e., the error probability and
the error bar in the asymptotic framework. 
That is,
the above four advantages correspond to 
respective criteria.
Therefore, limiting distribution provides a unified framework for these criteria.
The first three criteria are familiar in statistics, and the forth criterion is familiar in computer science.

In the present paper, we analyze the limiting distribution in the phase estimation systematically, 
and show that the limiting distribution is expressed by Fourier transform of a square integrable function on the closed interval $[-1,1]$, 
which approximately gives the input state in the asymptotic setting.

In the realistic setting,
the optimization of the forth criterion is more appropriate than that of the first criterion, i.e., 
the optimization of the mean square error or the average fidelity.
However, 
the estimator only with several special input states
were treated in the forth formulation, 
and its optimization was not discussed while
the optimization concerning the mean square error and the average
fidelity were done in the existing researches \cite{Buzek,Adaptive phase estimation}.

In the statistics, in order to treat this problem, they consider interval
estimation, in which, our inference is given as an interval. 
Indeed, 
there are two formulations in the statistics; one is the point estimation, in which, our estimate is given as only one point, and the other is the interval estimation.
In the point estimation, 
it is not easy to guarantee the quality of our estimate
because the estimated value always has statistical fluctuation. 
In order to resolve this problem, 
in the interval estimation,
for given data and confidence coefficient, 
our estimate is given as a interval, which is called a confidence interval. 
In this formulation, smaller width of confidence interval is better. 
As is mentioned in Section \ref{s2}, when number
$n$ of data is sufficiently large, 
the confidence interval can be provided
by using limiting distribution \cite{Statistics}.

Further, 
we analyze the variance and the tail probability of 
limiting distribution.
As our result,
concerning the limiting distribution,
we prove that the limiting distribution minimizing the asymptotic variance does
not minimize the tail probability. 
It is also shown that the limiting distribution minimizing the tail probability
depends on the width of interval. 
The definition of the tail probability depends on the width of interval. 
Such an optimal input state is given by a prolate spheroidal
wave function\cite{SlepianPollak} which often appears in wavelet
or time-limited Fourier analysis. This function is a solution of
the linear differential equation \cite{Slepian}: \[
\frac{d}{dx}(1-x^{2})\frac{df}{dx}+(\xi(R)-R^{2}x^{2})f=0.\]
Originally, prolate spheroidal wave functions appears in analysis on Helmholtz equation in electromagnetics\cite{electromagnetics} or determination of laser mode \cite{laser}.
Employing this wave function,
Slepian and Pollak \cite{SlepianPollak} 
extended Shannon's sampling theorem the case where the 
time-interval is limited as well as the bandwidth
while the original Shannon's sampling theorem treats the bandwidth-limited case.

Further, by using these facts, we study optimal interval estimation.
We provide the estimation protocol minimizing 
the width of confidence interval that assures the given confidence coefficient. 
In this case, the optimal estimation protocol depends
on a given confidence coefficient. That means, we must prepare input
state properly depending on the confidence coefficient.

The paper is organized as follows. 
In Section \ref{s2}, 
the formulation of the phase estimation is given and the limiting distribution is 
introduced with explanation of its meaning.
In Section \ref{s2-5}, we 
clarify the relation between the limiting distribution and Fourier transform. 
In Section \ref{s3}, 
we analyze the variance of the
limiting distribution. This problem is reduced to find the minimum
eigenvalue of a operator in the Dirichlet problem. 
In section \ref{s4},
the tail probability of the limiting distribution is discussed.
It is shown that the limiting distribution minimizing the variance does not provide a small tail probability.
In section \ref{s5},
we treat interval estimation problem. 
This problem can be analyzed by a prolate spheroidal wave function
and the eigenvalue of the defining differential equation.
In section \ref{s6},
the phase estimation with a single copy is discussed in the continuous system.
The discussions in Sections \ref{s3} and \ref{s4} can be applied to this formulation 
under the deterministic energy constraint.
In section \ref{s7}, we shortly note on the asymptotic Cram\'er-Rao lower-bound.

\section{Limiting distribution}\label{s2}

Let us consider the estimation problem 
of the unknown phase shift $\theta$
with an $n$-fold unitary evolution $V_\theta ^{\otimes n}$ of 
the unitary
$V_\theta:=
\left[\begin{array}{cc}
e^{i\frac{\theta}{2}} & 0\\
0 & e^{-i\frac{\theta}{2}}\end{array}\right]$,
in which our estimating protocol is given 
by a combination of an appropriate input state 
$\left|\phi_{0}\right\rangle $ 
and a suitable measurement $M$ (See Fig. \ref{fig-10-7}).

\begin{figure}[htbp]
\begin{center}
$\left|\psi\right\rangle \rightarrow$\fbox{$V_{\theta}^{\otimes n}$}$\rightarrow\left|\psi_{\theta}\right\rangle \rightarrow$\fbox{$M$}$\rightarrow P_{\theta,\psi}^{M}$
\end{center}
\caption{Our estimation scheme}
\label{fig-10-7}
\end{figure}

As is mentioned later, 
this formulation essentially contains 
the most general framework with $n$ applications of the unknown unitary.
In the following discussion, 
the true parameter is described by 
$\theta$ and our estimate is by $\hat\theta$.

As is shown in Appendix \ref{A2},
this problem is equivalent to estimate
the parameter $\theta$ of the unitary operation: 
\begin{align}
U_{\theta}=\sum_{k=0}^{n}e^{i(k-\frac{n}{2})\theta}\left|k\right\rangle \left\langle k\right|.
\label{12}
\end{align}
Our scheme for estimating $\theta$ is as follows. First, prepare
an input state $\left|\phi_{0}\right\rangle =\sum_{k=0}^{n}a_{k}\left|k\right\rangle $
where the coefficients $\vec{a}=\{a_{k}\}_{k=0}^{n}$ satisfy the normalizing condition $\sum_{k}\left|a_{k}\right|^{2}=1$. 
Second, evolve the input state $\left|\phi_{0}\right\rangle $ by
the unitary evolution $U_{\theta}$. And the last, perform a measurement
described by a POVM $M=M(\hat{\theta})d\hat{\theta}$. Then, the estimate
$\hat{\theta}$ obeys the probability distribution \[
\mathrm{P}_{\theta,\vec{a}}^{M}(\hat{\theta}):=\left\langle \phi_{\theta}\right|M(\hat{\theta})\left|\phi_{\theta}\right\rangle ,\]
 where \[
\left|\phi_{\theta}\right\rangle :=U_{\theta}\left|\phi_{0}\right\rangle =\sum_{k=0}^{n}a_{k}e^{i(k-\frac{n}{2})\theta}\left|k\right\rangle .\]

When our error function is given by
$R(\theta,\hat{\theta})$, we
optimize the mean error $D_{\theta}(M,\vec{a}):=\int_{0}^{2\pi}R(\theta,\hat{\theta})\mathrm{P}_{\theta,\vec{a}}^{M}(\hat{\theta})d\hat{\theta}$.
We only consider the covariant framework, i.e.,
the error function $R(\theta,\hat{\theta})$
is assumed to be given by a function of the difference 
$(\theta-\hat\theta) \mod 2\pi \mathbb{Z}$.
For example, when we focus on the gate fidelity 
$|\frac{\Tr V_\theta V_{\hat\theta}^{-1}}{2}|^2
=\cos^2 \frac{\theta-\hat\theta}{2}$,
the error $1- |\frac{\Tr V_\theta V_{\hat\theta}^{-1}}{2}|^2
=\sin^2 \frac{\theta-\hat\theta}{2}$ satisfies the covariant condition.

Then,
our measurement may be restricted into a group covariant measurement \begin{eqnarray}
M_{|t\rangle}(\mathrm{d}\hat{\theta}):=U_{\hat{\theta}}|t\rangle\langle t|U_{\hat{\theta}}^{\dagger}\frac{\mathrm{d}\hat{\theta}}{2\pi},\label{19-1}\end{eqnarray}
 where 
\begin{eqnarray}
|t\rangle=\sum_{k=0}^{n}e^{i\xi_{k}}\left|k\right\rangle .\label{19-2}
\end{eqnarray}
 This is because the minimum of the Bayesian average value $\min_{M}\frac{1}{2\pi}\int_{0}^{2\pi}D_{\theta}(M,\vec{a})d\theta$
under the invariant prior and the mini-max value $\min_{M}\max_{\theta}D_{\theta}(M,\vec{a})$
can be attained by the same group covariant measurement\cite{Holevo1}.
Therefore, we restrict our measurement to covariant measurements in the following discussion.
Then, our protocol is described by the pair of the coefficient of the input state $\vec{a}$
and the vector $|t\rangle $ given in (\ref{19-2}).

Further, without loss of generality, we can restrict our protocol to the pair of $(\vec{a},|t_0\rangle)$ as follows, where 
\begin{eqnarray}
|t_0\rangle=\sum_{k=0}^{n}
\left|k\right\rangle .\label{10-1}
\end{eqnarray}
For 
any protocol $(\vec{a},|t\rangle)$, we define $\vec{a}'=\{a_k\}$ by
\begin{align}
a_k' :=a_k  e^{-i\xi_{k}}.
\end{align}
Then, as is explained below, 
the protocol $(\vec{a}',|t_0\rangle)$ has the same performance as
the protocol $(\vec{a},|t_0\rangle)$.
\begin{align*}
& \mathrm{P}_{\theta,\vec{a}}^{M_{|t\rangle}}(\hat{\theta})
  = |\langle\phi_{\theta}|U_{\hat{\theta}}|t\rangle|^{2}
  = 
\left|\sum_{k=0}^{n}\overline{a_{k}}\left\langle k|U_{\hat{\theta}-\theta}t\right\rangle \right|^{2}
\nonumber \\
  = &
\left|\sum_{k=0}^{n}\overline{a_{k}}e^{i\xi_{k}}e^{i(k-\frac{n}{2})(\hat{\theta}-\theta)}\right|^{2}
  = 
\left|\sum_{k=0}^{n}\overline{a_{k}}e^{i\xi_{k}}e^{i(k-\frac{n}{2})(\hat{\theta}-\theta)}\right|^{2}
\nonumber \\
  = &
\left|\sum_{k=0}^{n}\overline{a_{k}'}e^{i(k-\frac{n}{2})(\hat{\theta}-\theta)}\right|^{2}
= \mathrm{P}_{\theta,\vec{a}'}^{M_{|t_0\rangle}}(\hat{\theta})
.\nonumber 
\end{align*}
Therefore, the choice of our protocol is essentially given by the choice of input state.
Dam et al \cite{Dam} proved this argument in a more general framework as follows.
When the number of application is $n$,
any protocol can be simulated by the above formulation.
That is, any adaptive application of the unknown unitary $V_\theta$ can be simulated by the $n$-fold unitary evolution $V_\theta ^{\otimes n}$ under the above error function.

The main target of the present paper is analyzing the asymptotic behavior of output distribution
for the sequence of input states ${\cal M}:=\{\vec{a}^n\}$.
For this purpose, we treat the distribution concerning the parameter
$z_n=\frac{n(\hat{\theta}_n-\theta)}{2}$ because the estimate
$\hat{\theta}_n$ approaches the true parameter $\theta$ with the order
$\frac{1}{n}$ when an appropriate measurement and an appropriate input state are used.
When the random variable $z_n$ converges to a random variable $z$ in probability,
the distribution $P({\cal M})$ of $z$ is called the limiting distribution of 
the sequence of input states ${\cal M}$.
In the case of state estimation including the classical case, 
if we apply a suitable estimator, 
the limiting distribution is the Gaussian distribution under a suitable regularity condition.
In the classical case, 
more precisely,
the asymptotic sufficient statistics for the given parameter 
obeys this Gaussian distribution.
That is, 
any estimator is given as a function of the statistics obeying the Gaussian distribution, asymptotically.
This Gaussian distribution is characterized only by the variance.
Even in the quantum case of state estimation,
the estimation problem can be reduced in that of quantum Gaussian states family in the asymptotic sense\cite{GK,GJK,GJ}.
In particular, 
if we treat the estimation of one-parameter model,
we obtain the same conclusion as is in the classical case.
Hence, it is sufficient to evaluate the variance for considering the limiting distribution.
That is, there is no variety concerning the limiting distribution of  
the state estimation because
the limiting distribution is uniquely determined as the Gaussian distribution.
The main problem in the present paper is, on the other hand, considering
whether there exist a variety concerning the limiting distribution of  
the phase estimation.

When the cost function $R(\theta,\hat\theta)$ has the form
$R(\theta,\hat\theta)\cong c(\theta-\hat\theta)^2+ o((\theta-\hat\theta)^2)$,
the average error behaves as
$D_\theta(M_{|t_0\rangle},\vec{a}_n)
 \cong \frac{c}{n^2} V$, where $V$ is the variance of the limiting distribution.
For example, $R(\theta,\hat\theta)= \sin^2 \frac{\theta-\hat\theta}{2}$,
the constant $c$ is $\frac{1}{4}$.
Hence, the analysis on the limiting distribution yields 
the asymptotic analysis on average gate fidelity.
Further, the analysis on the limiting distribution provides 
the asymptotic analysis on the phase estimation from another aspect.
For example, the tail probability of the sequence of input states ${\cal M}$
can be described as follows:
\begin{eqnarray}
\mathrm{P}_{\theta,\vec{a}^n}^{M_{|t_0\rangle}}
\{
|\hat{\theta}_n-\theta|> \frac{A}{n}
\}
\to 
P({\cal M})\{|z|> A\}.
\end{eqnarray}
Using this relation,
we can evaluate the required number of applications of the unknown unitary gate
for the given allowable error width $B$ and the given allowable error probability $\epsilon$ as follows.
First, we choose $A$ by $P({\cal M})\{|z|> A\}=\epsilon$.
Next, we choose $n$ by $\frac{A}{n}=B$, i.e., $n=\frac{A}{B}$.
Thus, the required number of applications is equal to $\frac{A}{B}$
if we use the sequence of input states ${\cal M}$.
The above discussions clarify that
the analysis on the limiting distribution yields
various types of asymptotic analysis on phase estimation.

Hence, in the present paper, for a deeper and unified asymptotic analysis on phase estimation,
we analyze the limiting distribution of the sequence of input states ${\cal M}$.

\section{Relation with square integrable functions}\label{s2-5}
In this section, we give a remarkable relation between limiting distributions
and square integrable functions on $[-1,1]$.

\begin{thm}
For any sequence of input states ${\cal M}$ having the limiting distribution,
there exists a function $f \in L^2([-1,1])$ 
satisfying the normalizing condition 
$\int_{-\infty}^{\infty}\left|f(x)\right|^{2}\mathrm{d}x=1$
and 
\begin{align}
P({\cal M})= \mathrm{P}^f,
\label{10-7-1}
\end{align}
where
$\mathrm{P}^{f}(\mathrm{d}z)=\left|F(f)(y)\right|^{2}\mathrm{d}z
$ and $F(f)$ is the Fourier transform on $L^2(\mathbb{R})$ of $f$, i.e., 
$F(f)(y):=\frac{1}{\sqrt{2\pi}}\int_{-\infty}^{\infty}f(x)e^{ixy}\mathrm{d}x$.

Conversely, 
for a function $f \in L^2([-1,1])$ with the normalizing condition
there exists a sequence of input states ${\cal M}$ 
satisfying the condition (\ref{10-7-1}).
\end{thm}

Due to this relation,
we can reduce the analysis on limiting distributions to 
the analysis on wave functions on the interval $[-1,1]$.
That is, our problem results in Fourier analysis on the interval $[-1,1]$.

\begin{proof}
As the first step of the proof, 
we construct 
a function $f \in L^2([-1,1])$.
for a given sequence of input states ${\cal M}:=\{ \vec{a}^n\}$.
For this purpose, we define
a function $f\in L^2([-1,1])$ by 
\begin{align}
f_n(x):=\overline{a_{k}}\sqrt{\frac{n}{2}}\label{10-7-4}
\end{align}
 for $x\in(x_{k}-\frac{1}{n+1},x_{k}+\frac{1}{n+1}]$,
where $x_{k}:=\frac{2k-n}{n+1}$.
In the following, the set of the above $L^2$ functions is denoted by $L^2_n$.
The parameter $z_n=\frac{n(\hat{\theta}_n-\theta)}{2}$
can be replaced by 
the parameter $y_n=\frac{(n+1)(\hat{\theta}_n-\theta)}{2}$
because the ratio $\frac{y_n}{z_n} \to 1$.
Since 
\begin{eqnarray*}
 &  & \int_{x_{k}-\frac{1}{n+1}}^{x_{k}+\frac{1}{n+1}}f_n(x)e^{ixy_n}\mathrm{d}x\\
 & = & \overline{a_{k}^n}
\sqrt{\frac{n+1}{2}}\frac{2}{y_n}\sin\frac{y_n}{n+1},\end{eqnarray*}
we have 
\begin{align}
   &\mathrm{P}_{\theta,\vec{a}^n}^{M_{|t_0\rangle}}(\hat{\theta}_n)\frac{1}{2\pi}\frac{d\hat{\theta}_n}{dy_n}\mathrm{d}y_n \nonumber \\
  = & \frac{1}{\pi (n+1)}\left|\sum_{k=0}^{n}\overline{a_{k}^n}e^{i(k-\frac{n}{2})(\hat{\theta}_n-\theta)}\right|^{2}\mathrm{d}y_n\nonumber \\
  = & \frac{1}{2\pi}\left|\sum_{k=0}^{n}\sqrt{\frac{2}{n+1}}\overline{a_{k}^n}
e^{ix_{k}y_n}\right|^{2}\mathrm{d}y_n\nonumber \\
  = & \frac{1}{2\pi}\left|\int_{-1}^{1}f_n(x)e^{ixy_n}\mathrm{d}x\frac{\frac{y_n}{n+1}}{\sin\frac{y_n}{n+1}}\right|^{2}\mathrm{d}y_n.\label{10-7-2}
\end{align}
When $f_n$ goes to a function $f \in L^2([-1,1])$,
since 
\begin{align}
\frac{\frac{y}{n+1}}{\sin\frac{y}{n+1}} \to 1 \label{10-7-3} 
\end{align}
as $n$ goes to infinity,
the distribution
of the normalized outcome $y=\frac{(n+1)(\hat{\theta}_n-\theta)}{2}(=y_n)$ 
convergences to the distribution $\mathrm{P}^{f}(\mathrm{d}y)$. 
If $f_n$ does not converge, 
by replacing the sequence $\{f_n\}$ by a converging subsequence $\{f_{n_k}\}$, i.e.,
$f_{n_k} \to f$,
we can show that 
the distribution of the normalized outcome converges 
the distribution $\mathrm{P}^{f}(\mathrm{d}y)$. 

Next, we prove the opposite argument.
For a given function $f \in L^2([-1,1])$ satisfying the normalizing condition,
we construct 
a sequence of input states satisfying (\ref{10-7-1}).
There exists a sequence of functions $f_n \in L^2_n$ such that $f_n \to f$.
Then, we define the coefficient $a_k^{(n)}$ by
$f_n(x_k)=\overline{a_{k}^{(n)}}\sqrt{\frac{n}{2}}$.
Due to (\ref{10-7-2}) and (\ref{10-7-3}), 
the sequence of input states ${\cal M}_f:=\{\vec{a}^{(n)}\}$
satisfies (\ref{10-7-1}).
\end{proof}

For example, when the input state is $\sum_{k=0}^{n}\sqrt{\frac{1}{n+1}}|k\rangle$,
the function $f$ is the constant $\sqrt{\frac{1}{2}}$. Since its
Fourier transformation is given by $\frac{1}{\sqrt{2\pi}}\frac{\sin y}{y}$,
the limiting distribution is described as \[
\mathrm{P}^{f}(\mathrm{d}y)=\left(\frac{\sin y}{y}\right)^{2}\frac{\mathrm{d}y}{2\pi}.\]

\section{Variance of Limiting Distribution}\label{s3}

In the previous section, we have shown that limiting distributions
$\mathrm{P}^{f}$ of outcomes are acquired through Fourier transforms
of wave functions $f\in L^2([-1,1])$ they correspond to coefficients
$a_{k}^{(n)}$ of input states. In this section, let us seek the 
input state minimizing the variance by utilizing this fact.
As is mentioned in Section \ref{s2}, 
optimizing the first-order coefficient of the variance is
equivalent with minimizing the variance 
$V(f):=\int_{-\infty}^{\infty}y^{2}\mathrm{P}^{f}(\mathrm{d}y)
$ over all functions $f\in L^2([-1,1])$.

Define the multiplication operator $Q$ and the momentum operator
$P=-i\frac{\mathrm{d}}{\mathrm{d}x}$ on $L^{2}(\mathbb{R})$. Then,
a function $f\in L^2([-1,1])$ satisfies that \[
V(f)=\left\langle f\right|F^{\dagger}Q^{2}F\left|f\right\rangle =\left\langle f\right|P^{2}\left|f\right\rangle .\]
 Under the natural embedding from $L^2([-1,1])$ to $L^{2}(\mathbb{R})$,
the minimum value of $V(f)$ is given by 
\[
\min_{f\in L^2([-1,1])}
\left\langle f\right|P^{2}\left|f\right\rangle ,\]
 which is nothing but the Dirichlet problem because the restriction of the operator $P^{2}$
on $L^{2}([-1,1])$ is equivalent with the square of the operator
$-i\frac{\mathrm{d}}{\mathrm{d}x}$ on 
\[
\{f\in L^{2}([-1,1])\cap C^{1}([-1,1])|f(-1)=f(1)=0\} .\]
That is, the problem is thus reduced to find the minimum eigenvalue
of the operator $P^{2}$. Its eigenvalues are $\pi^{2}m^{2}(m=1,2,\dots)$
and corresponding eigenfunctions are $\phi_{m}(x)=2\sqrt{2}\sin\pi m(\frac{x+1}{2})/C_m$
\cite{Coddington} where $C_m$ is a normalizing constant. 

Note that a careful treatment is required for a function $f\in L^{2}([-1,1])$
when the Dirichlet condition $f(-1)=f(1)=0$ does not hold. In this case, 
the function $f$ has a discontinuity at $1$ or $-1$ as an element of $L^{2}(\mathbb{R})$. 
Hence, the variance $\langle f|P^{2}|f\rangle$ is infinity. 
For example, in the case of $f=\sqrt{\frac{1}{2}}$, the
variance $\int_{-\infty}^{\infty}\sin^{2}y\frac{dy}{2\pi}$ diverges.
In this case, the limiting distribution is obtained with the order
$\frac{1}{n^2}$ 
\cite{CEMM} while the mean square
error goes to zero only the order $\frac{1}{n}$, i.e., we have a
quadratic speedup concerning limiting distribution, but no quadratic
speedup concerning mean square error. This fact is closely related
with the divergence of the integral $\int_{-\infty}^{\infty}\sin^{2}y\frac{dy}{2\pi}$.

\section{Tail Probability of Limiting Distribution}\label{s4}
In the asymptotic statistics, the behavior of tail probability of the  
limiting distribution
is one of the most important properties \cite{tail}
because it provides the performances of 
interval estimation and the powers of the one(or two)-side test.
Thus, we consider the tail probability of limiting distribution $ 
\mathrm{P}^{\phi_{m}}$.

In the statistics, the tail probability of the limiting distribution
is often discussed \cite{tail}. 
So, we consider the tail probability of limiting distribution $\mathrm{P}^{\phi_{m}}$.
In the i.i.d case, the minimum tail probability and minimum variance among limiting distributions
can be realized by the same Gaussian distribution.
However, in our setting, the Gaussian distribution does not minimize the variance.
Hence, it is not clear whether the minimum tail probability can be attained by the same distribution as the minimizing the variance.

Corresponding limiting distributions are acquired
through the Fourier transforms:\[
\mathrm{P}^{\phi_{m}}(y)=\left|F(\phi_{m})(y)\right|^{2}.\]
 Since
\begin{eqnarray*}
 &  & \int_{-1}^{1}C_m \phi_{m}(x)e^{ixy}\mathrm{d}x\\
 & = & \frac{1}{\sqrt{2}i}\int_{0}^{1}e^{i(y+m\pi)x}-e^{i(y-m\pi)x}\mathrm{d}x\\
 & = & \frac{1}{\sqrt{2}i}\left\{ \left[
\frac{e^{i(y+m\pi)x}}{i(y+m\pi)}
\right]_{x=0}^{1}-
\left[\frac{e^{i(y-m\pi)x}}{i(y-m\pi)}
\right]_{x=0}^{1}\right\} \\
 & = & \frac{1}{\sqrt{2}}(1-(-1)^{m}e^{iy})\left(\frac{1}{y+m\pi}-\frac{1}{y-m\pi}\right)\\
 & = & \frac{1}{\sqrt{2}}(-1+(-1)^{m}e^{iy})\frac{2m\pi}{y^{2}-m^{2}\pi^{2}},
\end{eqnarray*}
 limiting distributions are 
\[
\mathrm{P}^{\phi_{m}}(y)=\left|F(\phi_{m})(y)\right|^{2}=
\frac{2 m^{2}\pi (1-(-1)^{m}\cos y)
}{(y^{2}-m^{2}\pi^{2})^{2} C_m}.\]
 Thus, the tail probability of $\mathrm{P}^{\phi_{m}}$ decreases with
the order $O(y^{-4})$.
In order to improve the tail probability, 
we focus on the well-known fact that the Fourier transform
of a rapidly decreasing function is also a rapidly decreasing function.
In our problem, 
the support of the original wave function $f$ 
is included in $[-1,1]$.
Under this condition,
$f$ is a rapidly decreasing function
if and only if
$f$ is smooth function.
Note that a rapidly decreasing function does not decrease `suddenly'.
That is, the smoothness is an essential requirement.
For example, the function $\phi_m$ is not smooth at $-1$ and $1$.
In the following,
we construct a rapidly
decreasing wave function $f$ whose support is included in $[-1,1]$.
In this construction, the smoothing at $-1$ and $1$ is essential.

\begin{figure}[htbp]
\begin{center}
\includegraphics[width=8cm]{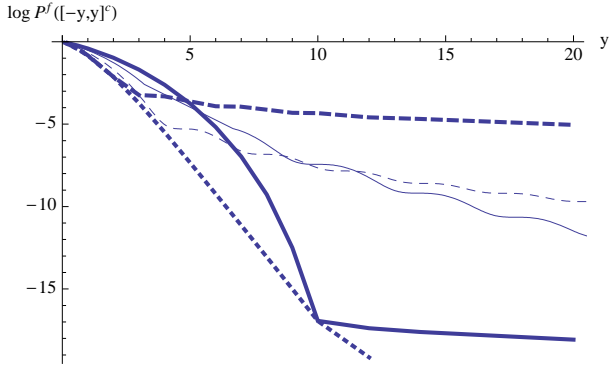}
\end{center}
\caption{
Logarithm of tail probabilities
$\log \mathrm{P}^{\phi_1}([-y,y]^c)$ (thin dashed),
$\log \mathrm{P}^{g_3}([-y,y]^c)$ (thin solid),
$\log \mathrm{P}^{\psi_2}([-y,y]^c)$ (thick dashed), 
$\log \mathrm{P}^{\psi_{10}}([-y,y]^c)$ (thick solid),
and 
Logarithm of
the minimum tail probability 
$\log \min_{f\in L^2([-1,1])}\mathrm{P}^{f}([-y,y]^{c})$
(thick dotted).}
\label{fig:3}
\end{figure}

First, functions $g_0$, $g_{1}$, and $g_{2}$ are defined by 
\begin{eqnarray*}
g_0(x) & := & \left\{ \begin{array}{cl}
\frac{2\exp(-1/x)}{\sqrt{x}} & \mathrm{if}~x>0\\
0 & \mathrm{otherwise,}\end{array}\right.\\
g_{1}(x) & := & g(x+1),\\
g_{2}(x) & := & g(1-x).
\end{eqnarray*}
Using these functions, we define a rapidly decreasing $g_3$ whose
support is included in $[-1,1]$ by 
\[
g_3(x)=g_{1}(x)g_{2}(x)/C\]
where $C$ is the normalizing constant.
As is checked numerically (See Fig. \ref{fig:3}.),
this function improves the tail probability.

Now, we analyze the decreasing speed on the tail probability of 
$\mathrm{P}^{g_3}([-R,R]^c)$.
Their Fourier transformations are 
\begin{eqnarray*}
F(g_0)(y) & = & \frac{1}{\sqrt{2}}\frac{\exp(-\sqrt{2|y|})}{\sqrt{|y|}}\exp\sgn(y)i(\sqrt{2|y|}+\frac{\pi}{4}),\\
F(g_{1})(y) & = & e^{-iy}F(g_0)(y),\\
F(g_{2})(y) & = & -e^{-iy}F(g_0)(-y),\\
F(g_3)(y) & = & \frac{1}{C}\sqrt{2\pi}F(g_{1})*F(g_{2})(y),\end{eqnarray*}
where $F(g_{1})*F(g_{2})$ is the 
convolution of $F(g_{1})$ and $F(g_{2})$.
When $y$ is sufficiently large, $|F(g_{i})(y)|^{2}\cong e^{-2\sqrt{2|y|}}$ for $i=1,2$,
i.e., \[
\lim_{y\to\infty}\frac{-1}{\sqrt{|y|}}\log|F(g_{i})(y)|^{2}=2\sqrt{2}, \quad i=1,2.\]
Then, as is shown in Appendix \ref{A1}, 
we obtain
\begin{equation}
 \lim_{y\to\infty}\frac{-1}{\sqrt{|y|}}\log2\pi|F(g_{1})*F(g_{2})(y)|^{2} \ge 2\sqrt{2}. \label{proof}
\end{equation}

Therefore, there exists a function $f$ such that the tail probability
of $\mathrm{P}^{f}$ is exponentially small and the support is included
in $[-1,1]$. Note that, the above wave function $g_3$ does not minimize
the variance $V(f)$. This fact tells us that the input state minimizing
the variance is not optimal concerning the tail
probability of the limiting distribution. That is, the optimal input
state depends on the choice of the criterion.

Next, we consider the maximization of the probability $\mathrm{P}^{f}([-R,R])$.
For this purpose, we denote the natural projection $\Pi_{R}$ from $L^{2}(\mathbb{R})$
to $L^{2}([-R,R])$. By using the operator $F_{R}:=F^{\dagger}\Pi_{R}F$,
this probability has the form \[
\langle f|F_{R}|f\rangle.\]
 That is, our aim is the following maximization: 
\begin{align*}
\max_{f\in L^{2}([-1,1])}\langle f|F_{R}|f\rangle
&=\max_{g\in L^{2}(\mathbb{R})}\frac{\langle g|\Pi_{1}F_{R}\Pi_{1}|g\rangle}{\|g\|^{2}} \\
&=
\max_{g \in L^{2}(\mathbb{R})}
\frac{\|\Pi_{R}F \Pi_{1} g\|^2}{\|g\|^2}
.
\end{align*}
This problem is equivalent with the calculation of the maximum eigenvalue of $\Pi_{1}F_{R}\Pi_{1}$. 

Slepian and Pollak \cite{SlepianPollak} showed that 
the eigenfunction $\psi_{R}$
of $\Pi_{1}F_{R}\Pi_{1}$ associated with the maximum eigenvalue
is given as
a solution of the linear differential equation,
which is called prolate spheroidal wave function: 
\[
\frac{d}{dx}(1-x^{2})\frac{df}{dx}+(\xi(R)-R^{2}x^{2})f=0,\]
where
$\xi(R)$ is chosen depending on the minimum eigenvalue.

Slepian \cite{Slepian} showed that
the maximum eigenvalue $\lambda(R)$ 
of $\Pi_{1}F_{R}\Pi_{1}$
behaves as
\begin{equation}
1-\lambda(R)\cong4\sqrt{\pi R}e^{-2R}(1-\frac{3}{32R}+O(R^{-2})) \label{eq:lambda}
\end{equation}
when $R$ is sufficiently large.
The numerical calculation of this minimum probability 
$\min_{f\in L^2([-1,1])}\mathrm{P}^{f}([-R,R]^{c})$
is given in Fig. \ref{fig:3}.
Thus, 
the minimum probability $\min_{f\in L^2([-1,1])}\mathrm{P}^{f}([-R,R]^{c})$
can be evaluated as \[
\lim_{R\to\infty}\frac{-1}{R}\log\min_{f\in L^{2}([-1,1])}\mathrm{P}^{f}([-R,R]^{c})=2.\]
That is, the minimum tail probability $\min_{f\in L^{2}([-1,1])}\mathrm{P}^{f}([-R,R]^{c})$
goes to zero with exponential rate $2$. This optimal value is attained
when the input state is given by the eigenfunction $\psi_{R}$ of $P_{1}F_{R}\Pi_{1}$
associated with the maximum eigenvalue $\lambda(R)$.

\begin{figure}[htbp]
\begin{center}
\includegraphics[width=8cm]{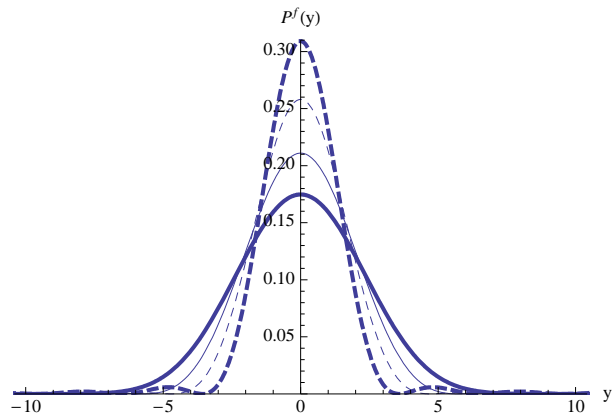}
\end{center}
\caption{
Probability density function minimizing variance $\mathrm{P}^{\phi_1}$ (thin dashed),
probability density function improving tail probability $\mathrm{P}^{g_3}$ (thin solid),
probability density function $\mathrm{P}^{\psi_{2}}$ (thick dashed), and
probability density function $\mathrm{P}^{\psi_{10}}$ (thick solid).
}
\label{fig:4}
\end{figure}

Now, we numerically compare 
the functions $\phi_1$, $g_3$, $\psi_{2}$, and $\psi_{10}$.
The density functions of the distributions
$\mathrm{P}^{\phi_1}$, $\mathrm{P}^{g_3}$, $\mathrm{P}^{\psi_{2}}$, $\mathrm{P}^{\psi_{10}}$ 
are plotted in Fig. \ref{fig:4}.
Their tail probabilities are plotted in Fig. \ref{fig:3}.
The tail probabilities
$\mathrm{P}^{\psi_2}([-y,y]^c)$ (thick dashed) and
$\mathrm{P}^{\psi_{10}}([-y,y]^c)$ (thick solid)
attain the minimum tail probability 
$\min_{f\in L^2([-1,1])}\mathrm{P}^{f}([-y,y]^{c})$ only at $2$ and $10$, respectively.
The distributions $\mathrm{P}^{\phi_1}$ and $\mathrm{P}^{\psi_{2}}$
concentrate in the range $[-2,2]$, however, their tail probabilities  
are not decreasing as rapidly as
those of the distributions $\mathrm{P}^{g_3}$ and $\mathrm{P}^{\psi_{10}}$.
This comparison indicates that 
the optimizations of the concentration and the tail probability are not compatible.
That is, the distributions of the Fourier transforms of
the functions $g_3$ and $\psi_{10}$ have a small tail probability (Fig. \ref{fig:4}).
These functions are smooth at $-1$ and $1$.
That means, we have checked that the smoothness is closely related to  
the tail probability.

\begin{figure}[htbp]
\begin{center}
\includegraphics[width=8cm]{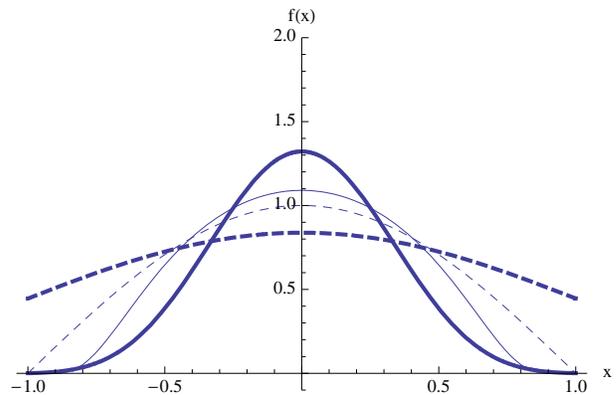}
\end{center}
\caption{
The wave function minimizing the variance $\phi_1$ (thin dashed),
the rapidly decreasing function $g_3$
whose support is included in $[-1,1]$ (thin solid),
prolate spheroidal wave function $\psi_{2}$ (thick dashed),
and prolate spheroidal wave function $\psi_{10}$ (thick solid).
} 
\label{fig:2}
\end{figure}

Next, we generalize this problem slightly, i.e., we maximize the probability
$\mathrm{P}^{f}([R_{1},R_{2}])$. In this case, the maximum value
coincides with that of $\mathrm{P}^{f}([(R_{1}-R_{2})/2,(R_{2}-R_{1})/2])$,
and its maximum is attained by the function $e^{ix(R_{1}+R_{2})/2}\psi_{(R_{2}-R_{1})/2}(x)$.

Since the function $R\mapsto\mathrm{P}^{f}([-R,R])$ is a strictly
monotone increasing function, the inverse function $\beta\mapsto R(\beta)$
is a strictly monotone increasing function. Thus, \[
\min_{f\in L^2([-1,1])}\min\{R|\mathrm{P}^{f}([-R,R])\ge\beta\}=R(\beta).\]
Further, the LHS coincides with \[
\min_{f\in L^2([-1,1])}\min\{R|\mathrm{P}^{f}([-R+a,R+a])\ge\beta\}\]
for any real number $a$.

\section{Interval estimation}\label{s5}

Now, we treat the phase estimation problem with the interval estimation.
In the interval estimation, given a confidence coefficient $\beta$,
we estimate the confidence interval $[L,U]$, which the unknown parameter
$\theta$ is guaranteed to belong to with the probability $\beta$.
Here, since our parameter space is the torus $\mathbb{R}/2\pi\mathbb{Z}$,
a careful treatment is required for the confidence interval $[L,U]$.
That is, for $L,U\in[0,2\pi)$, the confidence interval $[L,U]$ is
defined as a subset of $\mathbb{R}/2\pi\mathbb{Z}$ by

\[
[L,U]:=\left\{ \begin{array}{ll}
[L,U] & \hbox{if }L<U\\
{[L,U+2\pi]} & \hbox{otherwise,}\end{array}\right.\]
 and its width is defined \[
|[L,U]|:=\left\{ \begin{array}{ll}
U-L & \hbox{if }L<U\\
U+2\pi-L & \hbox{otherwise.}\end{array}\right.\]
In the interval estimation, the upper bound $U$ and the lower bound
$L$ of the interval are chosen from the outcome $\omega$ obeying
the distribution $\mathrm{P}_{\theta,\vec{a}}^{M}$. Since a smaller
width $|[L(\omega),U(\omega)]|$ is better, we minimize the width
$|[L(\omega),U(\omega)]|$ with the condition
$\mathrm{P}_{\theta,\vec{a}}^{M}\{\omega|\theta\in[L(\omega),U(\omega)]\}\ge\beta$ for any $\theta\in[0,2\pi)$. 
That is, we consider
\begin{widetext}
\begin{eqnarray}
 &  & \min_{U,L,M,\vec{a}}\left\{ \left.\max_{\omega}|[L(\omega),U(\omega)]|\right|\mathrm{P}_{\theta,\vec{a}}^{M}\{\omega|\theta\in[L(\omega),U(\omega)]\}\ge\beta,\quad\forall\theta\in[0,2\pi)\right\} \nonumber \\
 & = & \min_{U,L,M,\vec{a}}\max_{\theta\in[0,2\pi)}\left(\max_{\omega}\left\{ |[L(\omega),U(\omega)]|\left|\mathrm{P}_{\theta,\vec{a}}^{M}\{\omega|\theta\in[L(\omega),U(\omega)]\}\ge\beta\right.\right\} \right)\label{19-3}
\end{eqnarray}
 The value (\ref{19-3}) has a mini-max form of the cost $\max_{\omega}\left\{ |[L(\omega),U(\omega)]|\left|\mathrm{P}_{\theta,\vec{a}}^{M}\{\omega|\theta\in[L(\omega),U(\omega)]\}\ge\beta\right.\right\} $,
which has a covariant form. Thus, we can restrict our measurement
into covariant measurement (\ref{19-1}). Hence, our problem is reduced
as \[
\min_{t,\vec{a}}\left(\max_{\omega}\left\{ |[L(\omega),U(\omega)]|\left|\mathrm{P}_{\theta,\vec{a}}^{M}\{\omega|\theta\in[L(\omega),U(\omega)]\}\ge\beta\right.\right\} \right).\]
 However, since it is quite difficult to treat this optimization with
a finite $n$, we treat the following asymptotic setting as follows:
\begin{eqnarray*}
 &  & \lim_{n\to\infty}n\min_{t,\vec{a}}\left(\max_{\omega}\left\{ |[L(\omega),U(\omega)]|\left|
\mathrm{P}_{\theta,\vec{a}}^{M}\{\omega|\theta\in[L(\omega),U(\omega)]\}\ge\beta\right.\right\} 
\right)\\
 & = & \min_{f\in L^2([-1,1])}\max_{\omega}\left\{ |[L(\omega),U(\omega)]|\left|\mathrm{P}^{f}\left\{ \omega|\theta\in[L(\omega),U(\omega)]\right\} \ge\beta\right.\right\} \\
 & = & \min_{f\in L^2([-1,1])}\min\{2R|\mathrm{P}^{f}([-R,R])\ge\beta\}\\
 & = & 2R(\beta).
\end{eqnarray*}
\end{widetext}
This optimal value is attained when the input state constructed by
the wave function $\psi_{R(\beta)}$ and the measurement is given by
the covariant measurement (\ref{19-1}) with the vector $|t_0\rangle$.
That is, 
there exists a pair of functions $U$ and $L$
such that
$ |[L(\omega),U(\omega)]| \le 2 R(\beta)$
and $
\mathrm{P}_{\theta,\vec{a}}^{M}
\{\omega|\theta\in[L(\omega),U(\omega)]\}\ge\beta$.
The optimal input state depends on the choice of the confidence
coefficient $\beta$.

\section{Continuous case with single copy}\label{s6}

Let us consider the phase estimation in the continuous case with single copy,
in which 
by inputing the wave function $f$,
we estimate the parameter $\theta$ in 
a group-covariant model $\rho_{\theta}=e^{i\theta Q}\left|f\right>\left<f\right|e^{-i\theta Q}$
on the space $L^2(\mathbb{R})$.

It is known that when the shift-covariance condition is assumed for estimators,
our estimator is restricted into the measurement of the observable $P$\cite{Holevo3}.
Then, 
the outcome $\hat\theta$ obeys the distribution $\mathrm{P}^{f}$,
and the variance of the outcome is given by $\left<f| \Delta P^{2}|f\right> $,
 which is abbreviated by $\langle \Delta P^{2} \rangle$.

If we can input any wave function $f$, the variance can be reduced infinitesimally.
Hence, it is natural to assume a constraint for input wave function $f$.
Here, we assume that the potential is given as a monotone function of the absolute value $|Q|$.
While we often assume a constraint for average potential, we 
consider a deterministic condition for potential.
That is, 
the wave packet of $f$ is assumed to exist only in the region where the potential is less than a given constant.
In the following, for a simplicity for our analysis,
we assume that the input wave function belongs to $L^2([-1,1])$.
Hence, the discussion in Sections \ref{s3} and \ref{s4} can be applied to this problem.

Here, it is meaningful to consider the relation with the Cram\'er-Rao bound.
It is known in general that the Fisher information $J_{\theta}$ for
a group-covariant model $\rho_{\theta}=e^{i\theta Q}\left|f\right>\left<f\right|e^{-i\theta Q}$
is given by $J_{\theta}=\left<\Delta Q^{2}\right>$
because the symmetric logarithmic derivative (SLD) is given by $Q-\left<Q \right>$\cite{Holevo3}.

Since the operator $P$ has a commutation relation with $Q$, 
we
have the Heisenberg limit $\left<\Delta P^{2}\right>\left<\Delta Q^{2}\right>\geq1/4$, which 
is equivalent with the Cram\'er-Rao inequality:\[
\left<\Delta P^{2}\right>\geq\frac{1}{4}J_{\theta}^{-1}.\]
Especially, 
if and only if $f$ is a squeezed state
satisfying $\left<\Delta P^{2}\right>=c$ and $\left<\Delta Q^{2}\right>=c^{-1}$, 
the above inequality is achievable 
because its attainability is equivalent with that of
$\left<\Delta P^{2}\right>\left<\Delta Q^{2}\right>\geq1/4$.
Thus, if $f$ is not a squeezed state, the Cram\'er-Rao lower bound $\frac{1}{4}J_{\theta}^{-1}$
cannot be attained uniformly in the one-copy case.
As is shown in the next section, our asymptotic case is essentially equivalent to the above group-covariant
model under the restriction of $\mathrm{supp}f\subset[-1,1]$.

\section{Asymptotic Cram\'er-Rao Lower Bound}\label{s7}
Now, we consider the relation of our discussion with the Cram\'er-Rao lower bound
because the Cram\'er-Rao approach is often employed in the asymptotic estimation of 
the unknown unitary\cite{Imai}.
When we apply the sequence of protocols ${\cal M}:=\{\vec{a}^n\}$,
the phase estimation can be treated as the estimation problem in 
the state family $\{
|\phi_{\theta,n}\rangle \langle \phi_{\theta,n}| | \theta \in [0,2\pi] \}$, where
$\left|\phi_{\theta,n}\right\rangle =\sum_{k=0}^{n}a_{k}^{n}e^{ik\theta}\left|k\right\rangle $.
Let us calculate the SLD Fisher information.
From the group covariance of the output state, it suffices to calculate the SLD Fisher information at $\theta=0$. Let $\left|l_{\theta,n}\right\rangle :=(1-\left|\phi_{\theta,n}\right\rangle \left\langle \phi_{\theta,n}\right|)\frac{\partial}{\partial\theta}\left|\phi_{\theta,n}\right\rangle $.
The SLD Fisher information $J_{0,n}$ is given by 
$\frac{1}{4}J_{0,n}=\left\langle l_{0,n}\right|\left.l_{0,n}\right\rangle $.
\begin{eqnarray*}
\frac{J_{0,n}}{4} & = & \left\langle l_{0,n}\right|\left.l_{0,n}\right\rangle \\
 & = & \left\langle \phi_{0,n}'\right|\left.\phi_{0,n}'\right\rangle 
-\left|\left\langle \phi_{0,n}\right|\left.\phi_{0,n}'\right\rangle \right|^{2}\\
 & = & \sum_{k=0}^{n}k^{2}
\left|a_{k}^{n}\right|^{2}
-\left(\sum_{k=0}^{n}k\left|a_{k}^{n}\right|^{2}\right)^{2}
\end{eqnarray*}
 where $\left|\phi_{0,n}'\right\rangle 
=\left.\frac{\partial}{\partial\theta}\left|\phi_{\theta,n}\right\rangle \right|_{\theta=0}$.
Choosing a smooth function $f_n$ by (\ref{10-7-4}),
we have\[
\frac{J_{0,n}}{4(n+1)^{2}}=\sum_{k=0}^{n}
\left(x_{k}\right)^{2}\left|f_n(x_{k})\right|^{2}-
\left(\sum_{k=0}^{n}\left(x_{k}\right)\left|f_n(x_{k})\right|^{2}\right)^{2}.\]
When $f_n$ converges to $f$,
\begin{align*}
\lim_{n\rightarrow\infty}\frac{J_{0,n}}{4(n+1)^{2}}
=&\int_{-1}^{1}x^{2}\left|f(x)\right|^{2}\mathrm{d}x-\left(\int_{-1}^{1}x\left|f(x)\right|^{2}\mathrm{d}x\right)^{2} \\
=&
\left\langle f\right|\Delta Q^{2}\left|f\right\rangle.
\end{align*}
Since the variance of the limiting distribution
is $\left\langle f\right|\Delta P^{2}\left|f\right\rangle$,
we obtain the limiting distribution version of the 
Cram\'er-Rao inequality as
\begin{align*}
\left\langle f\right|\Delta P^{2}\left|f\right\rangle
\ge 
\frac{1}{\lim_{n\rightarrow\infty}\frac{1}{4(n+1)^{2}}J_{0,n}}
=\frac{1}{\left\langle f\right|\Delta Q^{2}\left|f\right\rangle}.
\end{align*}
The equality holds if and only if
the wave function $f$ is a squeezed state.
However, since the support $f$ belongs to $[-1,1]$,
the equality of the above cannot be attained.
This fact indicates that the Cramer-Rao approach does not yield the attainable bound in 
the estimation of unitary action even in the asymptotic formulation, 
while this approach generally yields the attainable bound in the estimation of quantum state.
This point is the essential difference between the state estimation and the unitary estimation.

\section{Conclusion}\label{s8}

As a unified approach to the asymptotic analysis on the phase estimation,
we have treated the limiting distribution on the sequence of estimators
because 
we can recover various asymptotic performance of the estimation protocols from the limiting distribution.

As the first step, we have found a one-to-one correspondence between
a limiting distribution and a wave function on $L^2([-1,1])$.
That is, we have shown that 
any limiting distribution is given by the absolute square of the Fourier transform of a wave function $f\in  L^2([-1,1])$.
Due to this correspondence, it is sufficient to optimize the distribution
given as the square of the Fourier transform on $L^2([-1,1])$.

As the next step, the minimization of the variance has been treated among the above distributions
by treating the Dirichlet problem in the similar way as Buzek et al \cite{Buzek}.
We have also considered its tail probability.
In order to guarantee the small error probability out of the given interval,
the limiting distribution is better to be rapidly decreasing.
However, it has been clarified that the limiting distribution minimizing the variance 
is not rapidly decreasing.
In order to construct such a limiting distribution,
we employ a smoothing method so that 
we construct a rapidly decreasing function whose support is included in $[-1,1]$.
It has been numerically checked that this function improves the tail probability remarkably.

Further, the tail probability for a given interval
has been minimized among these limiting distribution by employing the Slepian and Pollak's analysis on signal processing\cite{SlepianPollak}.
The optimal limiting distribution depends on the width of this interval.
Using this optimization, we have treat the interval estimation in the asymptotic setting.

Next, we have treated the relation with the phase estimation in the continuous system with the one copy setting.
In this case, the Heisenberg's uncertainly relation is equivalent with 
Cram\'{e}r-Rao inequality.
Using this relation, we have obtained the condition for attainability of 
Cram\'{e}r-Rao inequality.
Further, we have applied this relation to the asymptotic analysis on the variance of the phase estimation.
Then, we have clarified that the Cram\'{e}r-Rao bound cannot be attained in our framework.

Throughout these discussions, it has been clarified that
the optimization of asymptotic phase estimation cannot be characterized
by a single parameter while 
this problem can be characterized by the single parameter, i.e., the variance,
in the state estimation of a single parameter model with a regularity condition
due to the asymptotic normality\cite{GK,GJK,GJ}.
This property is the biggest difference from the state estimation.

Indeed, a similar property can be expected in a general unitary estimation.
It is a future problem to investigate the limiting distribution in the estimation of
unitary operation in a more general case.

\section*{Acknowledgment}

The authors thank Professor Toshiyuki Sugawa, Professor Fumio Hiai,
and Professor Fuminori Sakaguchi for discussing Fourier analysis. 
The authors also thank to Professor Michele Mosca for discussion
about quantum circuits.

This research was partially supported by a Grant-in-Aid for Scientific
Research on Priority Area `Deepening and Expansion of Statistical
Mechanical Informatics (DEX-SMI)', no. 18079014. 

\appendix

\section{Elimination of multiplicity}\label{A2}
The unitary $V_\theta^{\otimes n}$
can be written as the form
\[
U_{\theta}'=
\sum_{k=0}^{n}\sum_{j=1}^{m_k}
e^{i(k-\frac{n}{2})\theta}
\left|k,j\right\rangle \left\langle k,j\right|\]
with the multiplicity $m_k= {n\choose k}$.
When the unitary $U_{\theta}'$ acts on 
the input state 
$
\sum_{k=0}^{n}\sum_{j=1}^{m_k}
a_{k,j}
\left|k,j\right\rangle
$,
the final state is given by $
\sum_{k=0}^{n}\sum_{j=1}^{m_k}
e^{i(k-\frac{n}{2})\theta}
a_{k,j}
\left|k,j\right\rangle
=
\sum_{k=0}^{n}
e^{i(k-\frac{n}{2})\theta}
(
\sum_{j=1}^{m_k}
a_{k,j}
\left|k,j\right\rangle)
=
\sum_{k=0}^{n}
e^{i(k-\frac{n}{2})\theta}
a_{k}
\left|k\right\rangle)
$
where $a_{k}
:=\sqrt{\sum_{j=1}^{m_k}
|a_{k,j}|^2}$ and $
\left|k\right\rangle:=
\frac{1}{a_{k}}
\sum_{j=1}^{m_k}
a_{k,j}
\left|k,j\right\rangle$.
Then, the estimation problem of $U_\theta'$
can be reduced in that of $U_\theta$ given in (\ref{12}).

\section{Proof of (\ref{proof})}\label{A1}
Now, we prove (\ref{proof}).
Assume that $y>0$.
For a given integer $N$,s
\begin{eqnarray*}
&  & 
\frac{\sqrt{2\pi}}{C}
|\int_{-\infty}^{\infty}
F(g_1)(y') F(g_2)(y-y') dy'| \\
& \le &
\sum_{k=1}^N 
\frac{\sqrt{2\pi}}{C}
\max_{y'\in [y\frac{k-1}{N},y\frac{k}{N}]}
|F(g_1)(y')|\cdot |F(g_2)(y-y')|
\frac{y}{N} \\
&& +
\frac{\sqrt{2\pi}}{C} \int_{-\infty}^0 
|F(g_1)(y')|\cdot | F(g_2)(y-y')| dy' \\
&& +\frac{\sqrt{2\pi}}{C} \int_{y}^{\infty}
|F(g_1)(y')|\cdot | F(g_2)(y-y')| dy'.
\end{eqnarray*}
Since $|F(g_1)(y')|$ is bounded,
\begin{align*}
&\frac{\sqrt{2\pi}}{C} \int_{-\infty}^0 
|F(g_1)(y')|\cdot | F(g_2)(y-y')| dy' \\
\cong &
O( \int_{-\infty}^0  | F(g_2)(y-y')| dy' )
\cong
O( e^{-\sqrt{2}\sqrt{y}}).
\end{align*}
Similarly, 
\begin{align*}
&& \frac{\sqrt{2\pi}}{C} \int_{y}^{\infty}
|F(g_1)(y')|\cdot | F(g_2)(y-y')| dy'
\cong 
O( e^{-\sqrt{2}\sqrt{y}}).
\end{align*}
Further,
\begin{align*}
& \frac{\sqrt{2\pi}}{C}
\max_{y'\in [y\frac{k-1}{N},y\frac{k}{N}]}
|F(g_1)(y')|\cdot |F(g_2)(y-y')|
\frac{y}{N} \\
\cong &
O( 
e^{-\sqrt{2}\sqrt{y \frac{k-1}{N}}}
\cdot
e^{-\sqrt{2}\sqrt{y-y \frac{k}{N}}})
\le
O( e^{-\sqrt{2}\sqrt{y (1-\frac{1}{N}})}).
\end{align*}
Therefore,
\begin{eqnarray*}
\frac{\sqrt{2\pi}}{C}
|\int_{-\infty}^{\infty}
F(g_1)(y') F(g_2)(y-y') dy'|
& \le &
O( e^{-\sqrt{2}\sqrt{y (1-\frac{1}{N}})}).
\end{eqnarray*}
Since $N$ is arbitrary,
\begin{eqnarray*}
\frac{\sqrt{2\pi}}{C}
|\int_{-\infty}^{\infty}
F(g_1)(y') F(g_2)(y-y') dy'|
& \le &
O( e^{-\sqrt{2}\sqrt{y }}).
\end{eqnarray*}
Taking the square, we obtain (\ref{proof}).

In the case of $y<0$, we can show (\ref{proof}) by replacing $y'$ by $-y'$.



\begin{thebibliography}{10}
\bibitem{CEMM} R. Cleve, A. Ekert, C. Macchiavello and M. Mosca,
{}``Quantum Algorithm Revisited,'' \textit{Proc. R. Soc. London,
Ser. A} 454, 339, 1998. 

\bibitem{Kitaev} A. Y. Kitaev, A. H. Shen and M. N. Vyalyi, \textit{Classical
and Quantum Computation, (Graduate Studies in Mathematics 47), }Americal
Mathematical Society, 2002.

\bibitem{Giovannetti}V. Giovannetti, S. Lloyd and L. Maccone, {}``Quantum-enhanced
measurements: beating the standard quantum limit,'' \textit{Science}
306, 1330-1336, 2004.

\bibitem{Buzek}V. Buzek, R. Derka and S. Massar, {}``Optimal Quantum
Clocks,'' \textit{Phys. Rev. Lett.} 82 (1999) 2207, quant-ph/9808042.

\bibitem{Hayashi}M. Hayashi, {}``Parallel Treatment of Estimation
of SU(2) and Phase Estimation,'' \textit{Phys. Lett. A} 354, 183-189,
2006.

\bibitem{Imai}H. Imai and A. Fujiwara, {}``Geometry of optimal estimation
scheme for SU(D) channels,'' \textit{J. Phys. A} 40, 4391-4400, 2007.

\bibitem{Heisenberg-limit}B. L. Higgins, D. W. Berry, S. D. Bartlett,
H. M. Wiseman and G. J. Pryde, {}``Entanglement-free Heisenberg-limited
phase estimation,'' \textit{Nature} 450, 393-396, 2007.

\bibitem{Kitaev2} A. Y. Kitaev, ``Quantum Computations: Algorithms and Error Correction,''
\textit{Russ. Math. Surv.} 52, 1191-1249, 1997.

\bibitem{HayashiMatsumoto}M. Hayashi and K. Matsumoto, {}``Asymptotic
performance of optimal state estimation in quantum two level system,''
\textit{quant-ph/0411073}.

\bibitem{GK}
M. Gu\c{t}\u{a} and J. Kahn, ``Local asymptotic normality for qubit states,''
{\em Phys. Rev. A}, {\bf 73}, 052108 (2006).

\bibitem{GJK}
M. Gu\c{t}\u{a}, B. Janssens, and J. Kahn,
``Optimal estimation of qubit states with continuous time measurements,''
{\em Commun. Math. Phys.}, {\bf 277}, 127-160 (2008).

\bibitem{GJ}
M. Gu\c{t}\u{a} and A. Jencova,
``Local asymptotic normality in quantum statistics,''
{\em Commun. Math. Phys.}, {\bf 276}, 341-379 (2007).


\bibitem{Adaptive phase estimation}D. Pope, H. M. Wiseman and N.
K. Langford, {}``Adaptive Phase estimation is more accurate than
nonadaptive phase estimation for continuous beams of light,'' \textit{Phys.
Rev. A} A70, 043812-1 - 043812-13 2004.

\bibitem{Statistics}R. J. Larsen and M. L. Marx, \textit{An Introduction
to Mathematical Statistics and Its Applications}, Pearson Education,
U.S., 2005.

\bibitem{SlepianPollak}D. Slepian and H. O. Pollak, {}``Prolate
spheroidal wave functions, Fourier analysis and uncertainty-I,''
\textit{Bell Syst. Tech. J.}, 40, 43-63, 1961.

\bibitem{Slepian}D. Slepian, {}``Some asymptotic expansions for
prolate spheroidal functions,'' \textit{J. Math. Phys.} 44, 99\textendash{}140,
1965.

\bibitem{electromagnetics}L. Li, M. Leong, T. Yeo and Y. Gan,
``Electromagnetic radiation from a prolate spheroidal antenna enclosed in a confocal spheroidal radome,''
\textit{IEEE Trans. on Antenn. and Propa.}, 50, 1525-1533, 2002.

\bibitem{laser}P. Nazmi, P. Kapadia and J. Dowden,
``A mathematical model of heat conduction in a prolate spheroidal coordinate system with applications to the theory of welding,''
\textit{J. Phys. D}, 26, 563-573, 1993.

\bibitem{Holevo1}A. S. Holevo, {}``Covariant Measurements and Uncertainty
Relations,''\textit{ Rep. Math. Phys.} 16, 385-400, 1979.

\bibitem{Coddington}E. A. Coddington and N. Levinson, \textit{Theory
of differential equations}, McGraw-Hill, New-York, 1955.

\bibitem{tail}F. P. Kelly (ed.),
\textit{Probability, Statistics and Optimization: A Tribute to Peter Whittle
(Wiley Series in Probability and Statistics)}, John Wiley \& Sons Inc., 1994.

\bibitem{Holevo2}A. S. Holevo, {}``Asymptotic estimation of shift
parameter of a quantum state,'' \textit{quant-ph/0307225}.

\bibitem{Holevo3}
 A. S. Holevo,
{\it Probabilistic and Statistical Aspects of Quantum Theory},
(North-Holland, Amsterdam, 1982);
Originally published in Russian (1980).

\bibitem{Dam}
W. van Dam, G. M. D'Ariano, A. Ekert, C. Macchiavello, and M. Mosca,
``Optimal quantum circuits for general phase estimation,''
\textit{quant-ph/0609160}.
\end{thebibliography}
\end{document}